\documentclass[a4paper,11pt]{article}
\pdfoutput=1 

\usepackage{jinstpub} 
\usepackage{subfig}
\usepackage{array}

\title{MARS-MD: rejection based image domain material decomposition.}

\author[a,b,c,d,1]{C.J. Bateman\note{Corresponding author.}}
\author[b]{, D. Knight}		
\author[g]{, B. Brandwacht}	
\author[c]{, J. Mc Mahon} 

\author[e]{, J. Healy}		
\author[a,b]{, R. Panta}		
\author[b]{, R. Aamir}		
\author[b]{, K. Rajendran}	
\author[b]{, M. Moghiseh}	
\author[b]{, M. Ramyar} 		

\author[a]{, D. Rundle}		
\author[a]{, J. Bennett}		
\author[a,b,f]{, N. de Ruiter}	
\author[a]{, D. Smithies}	
\author[a]{, S.T. Bell}		
\author[a]{, R. Doesburg}	

\author[a,f]{, A. Chernoglazov}	
\author[a,f]{, V.B.H. Mandalika}		
\author[a]{, M. Walsh}			
\author[a,b]{, M. Shamshad}		
\author[b]{, M. Anjomrouz}		
\author[b]{, A. Atharifard}		
\author[c]{, L. Vanden Broeke}   
\author[b,h]{, S. Bheesette}  	 	

\author[j]{, T. Kirkbride}	
\author[b]{, N.G. Anderson}	
\author[b,e]{, S.P. Gieseg}		
\author[b]{, T. Woodfield} 	
\author[b,c,i]{, P.F. Renaud}		
\author[a,b,h]{, A.P.H. Butler}		
\author[a,c,h]{, P.H. Butler}		

\affiliation[a]{MARS Bioimaging Ltd., 29a Clyde Rd, Christchurch, New Zealand.}
\affiliation[b]{Department of Radiology, University of Otago, Christchurch\\2 Riccarton Ave, Christchurch 8140, New Zealand.}
\affiliation[c]{Department of Physics and Astronomy, University of Canterbury,\\PO Box 69 133, Lincoln, Christchurch 7640, New Zealand.}
\affiliation[d]{Lincoln Agritech Limited, \\Private Bag 4800, Christchurch 8140, New Zealand.}
\affiliation[e]{Department of Biology, University of Canterbury, \\Private Bag 4800, Christchurch 8140, New Zealand.}
\affiliation[f]{HIT lab NZ, Christchurch, New Zealand.}
\affiliation[g]{University of Twente, Drienerlolaan 5, 7522NB Enschede, Netherlands.}
\affiliation[h]{European Organization for Nuclear Research (CERN), Geneva, Switzerland.}
\affiliation[i]{Department of Mathematics and Statistics, University of Canterbury, \\Private Bag 4800, Christchurch 8140, New Zealand.}
\affiliation[j]{Department of Nursing, Midwifery and Allied Health, \\Ara Institute of Canterbury, PO Box 540, Christchurch 8140, New Zealand}
\emailAdd{christopher.bateman@lincolnagritech.co.nz}

\abstract{%
This paper outlines image domain material decomposition algorithms that have been routinely used in MARS spectral CT systems. These algorithms (known collectively as MARS-MD) are based on a pragmatic heuristic for solving the under-determined problem where there are more materials than energy bins. This heuristic contains three parts: (1) splitting the problem into a number of possible sub-problems, each containing fewer materials; (2) solving each sub-problem; and (3) applying rejection criteria to eliminate all but one sub-problem's solution. An advantage of this process is that different constraints can be applied to each sub-problem if necessary. In addition, the result of this process is that solutions will be sparse in the material domain, which reduces crossover of signal between material images. Two algorithms based on this process are presented: the Segmentation variant, which uses segmented material classes to define each sub-problem; and the Angular Rejection variant, which defines the rejection criteria using the angle between reconstructed attenuation vectors. 
}


\begin{document}
\maketitle
\flushbottom


\section{Introduction}
Spectral CT is a variant of x-ray computed tomography (CT) which uses additional measurements of photon energy to enable identification and quantification of different materials simultaneously. Energy information can be obtained through a multitude of ways. Clinical dual-energy CT is typically performed using either dual-source \cite{Flohr2006}, kVp switching \cite{Kalender1986, Xu2009, Szczykutowicz2010}, or dual-layered detectors \cite{Carmi2005,Yao2014}. More recently there has been development of pre-clinical systems which utilise energy discriminating photon counting detectors to provide up to eight channels of energy information \cite{Bones2010,Persson2014,Sarno2016,Shikhaliev2015}. The MARS (Medipix All Resolution System) is a spectral CT system based on the Medipix3RX photon counting detector. This detector has eight energy bins which include one arbitrated, four charge summed, three single pixel counters. Information about these different counting modes can be found elsewhere \cite{Koenig2013}. Recent MARS systems typically only utilize the arbitration and/or charge-summed counters for spectral analysis due to their superior energy resolution, yielding 4-5 unique energy bins for each measurement.        
\newline \newline
The process of converting energy information into material information is known as either basis or material decomposition (MD). This was first demonstrated to be possible by Alzarez and Macovski through the conversion of dual-energy measurements into photo-electric and Compton cross-section components \cite{Alvarez1976}. Later this was extended to estimating effective density and atomic number \cite{White1977,Heismann2003}, and material specific densities \cite{Le2011, Roessl2007, Liu2009}. There are three different ways in which MD can be performed: pre-reconstruction, post-reconstruction, and (simultaneous) joint-reconstruction. Each of these have their own advantages and disadvantages. Pre- and joint- reconstruction allows for modelling of the x-ray beam poly-chromaticity, which in turn avoids beam hardening artefacts. To do this however requires use of non-linear optimization techniques which can be computationally intensive. Post-reconstruction MD has the advantage that it can be performed quickly, however it is susceptible to any beam hardening artefacts within the reconstructed data. The impact of this can be minimized by, using synchrotron x-ray sources, or applying beam-hardening correction techniques, and in some cases of measuring within narrow energy ranges.  
\newline \newline
One issue that is common to all three approaches is the linear dependencies between the attenuation distributions for materials commonly of interest in medical imaging. It was originally assumed that only two materials without K-edges could be decomposed due to the primary dependence on the photo-electric and Compton interactions in the diagnostic imaging range \cite{Firsching2004}. Bornefalk later showed that the attenuation for the set of elements $Z<20$ can theoretically be decomposed into four independent components \cite{Bornefalk2012}. Despite this, the decomposition of many materials simultaneously can still prove challenging if any of them are close in effective atomic number.      
\newline \newline 
This paper outlines two post-reconstruction MD algorithms that has been routinely used in MARS small animal scanners, and the heuristic they are based on to minimize the impact of this basis linear dependence issue. We also describe the image domain formulation of post-reconstruction MD problem. Use of the presented algorithms is demonstrated on three multi-material phantoms. In addition, references are given to several publications describe the use and results of MARS-MD in different pre-clinical applications.

\section{Image domain Material Decomposition}
MD in the image domain processes reconstructed energy bin volumes into material image volumes. Each voxel of the reconstructed image provides an estimate of the attenuation in that voxel for each energy bin. That voxel's attenuation for a given energy bin can be parametrized by a linear combination of basis functions
\begin{equation}
\mu(E) = a_1f_1(E) + a_2f_2(E) + \dots + a_N f_N(E)
\label{eqn:basis}
\end{equation}
Using this, the attenuation for all energy bins can be written as the matrix equation 
\begin{equation}
\textbf{b} = M\textbf{x}
\label{eqn:MD}
\end{equation}
where $\textbf{b} = \mu(\textbf{E})$ is a column vector containing the attenuation of the voxel in each energy bin; $M = [ f_1(\textbf{E}) \ f_2(\textbf{E}) \ \dots \ f_N(\textbf{E})]$ is a matrix whose columns containing values of the basis functions at the given energies; and $\textbf{b} = [a_1 \ a_2 \ \dots \ a_N]^T$ is the column vector containing the basis function coefficients. Here we use $\textbf{E}$ to represent the vector of energy bins with a slight abuse of notation. If each energy bin is capturing a monochromatic signal then the elements of $\textbf{E}$ would be just the energies of the corresponding x-rays. When working with polychromatic energy bins the elements of $\textbf{E}$ are more abstractly thought of as observable energy intervals. It is not uncommon for this to be defined using effective energies to simplify the description of each energy bin. The image domain MD problem is simply the inversion of this matrix equation ($\textbf{x} = M^{-1}\textbf{b}$ ) for every voxel in the volume.
\newline \newline
Numerous approaches have been taken for solving this problem, many of which are based on a linear least squares estimates \cite{Le2011, Ronaldson2011}. Variations include application of different constraints such as volume conservation \cite{Vinegar1987}, mass conservation \cite{Liu2009}, or non-negativity \cite{Zhang2014}; each voxel being treated either independent or dependent of its neighbours \cite{Clark2014}; and the representation of a voxel being restricted to a subset of basis functions \cite{Le2011,Bateman2013}.         
%
%
\section{MARS-MD Heuristic Process}
The MARS-MD algorithms are based on a simple heuristic which is designed to be adaptable to different situations. Both algorithms presented below are based on the same three steps: given a set of $N$ basis functions, (1) construct a number of feasible sub-problems from the basis set, (2) calculate the solution to every sub-problem, and (3) apply some rejection criteria to eliminate all but one sub-problem solution.
\newline \newline
When determining what set of sub-problems to use we first choose the maximum number $k$ of basis functions that will be allowed to describe a voxel (in practice this might be the maximum number of materials per voxel). We then consider all $k \leq N$ combinations of basis functions and further eliminate any that are non-physical and/or undesirable. A sub-problem is then formulated to find a solution to Eqn. \ref{eqn:MD} for each remaining sub-set of basis functions. Since each of these are calculated independently, different constraints can be applied to different sub-problems.      
\section{MARS-MD Algorithms}
There are two versions of the MARS-MD algorithm following this heuristic that have been used with MARS systems - the Segmentation variant and the Angular Rejection variant. These two algorithms have been known by different names in different places. In the MARS system automated image processing chain they are known as MARS-MD v1.0 and MARS-MD v1.1 respectively. Each of these methods only considers solutions with at most two materials per voxel. For each voxel, the MARS-MD algorithms typically decompose reconstructed effective linear attenuation onto a set of mass attenuation bases for different materials. The results of the algorithms are produced as estimated densities of the given basis materials. 
\subsection{Segmentation Variant}
\textbf{Sub-problem selection:} 
\newline 
This uses a segmentation procedure we have previously published \cite{Bateman2013} to categorize each voxel to contain either air, soft tissue, or dense material. Combinations of materials within a voxel are then restricted based on its class. Air voxels are considered to be empty and set to the zero solution. Soft tissue voxels are treated as containing a combination of lipid and water obeying volume conservation. Dense material voxels which contain either a single dense material, or a single dense material in combination with water.
\newline \newline
\textbf{Sub-problem solution:} 
\newline 
Eqn. \ref{eqn:MD} is solved for each basis combination using a non-negative linear least squares estimate. The volume constraint is applied to the lipid-water sub-problem as an additional equation included within the matrix equation. To do this the lipid and water bases used in the matrix equation need to be expressed as linear attenuation.  
\newline \newline
\textbf{Solution rejection:} 
\newline 
The linear least squares error $|| \textbf{b} - M\textbf{x}^* ||_2$ is calculated for each sub-problem solution $\textbf{x}^*$. The solution with the smallest error is selected as the final solution for the respective voxel.  
%
%
%
\subsection{Angular Rejection Variant}
\textbf{Sub-problem selection:} 
\newline
Each voxel is treated as possibly containing either a combination of lipid and water, a single dense material, or a single dense material in combination with water. These are the same basis combinations as used for the Segmentation variant, but without the additional layer of exclusion from the segmentation.  
\newline \newline
\textbf{Sub-problem solution:} 
\newline 
Eqn. \ref{eqn:MD} is solved for each basis combination using a non-negative linear least squares estimate.  
\newline \newline
\textbf{Solution rejection:} 
\newline 
An attenuation vector $\textbf{b}^*$ is calculated from each solution. These and the voxel's reconstructed attenuation estimate ($\textbf{b}$) are normalized by the corresponding attenuation for water (this is a pseudo-Hounsfield unit as outlined in \cite{Zainon2011}). The cosine angle between the normalized $\textbf{b}^*$ and $\textbf{b}$ vectors are calculated for each solution. The solution with the smallest cosine angle is selected as the final solution for the respective voxel.

\section{Algorithm Calibration}
\begin{figure}[h!]
	\centering
	\begin{tabular}{cc}
		\subfloat[]{\includegraphics[width=0.35\textwidth]{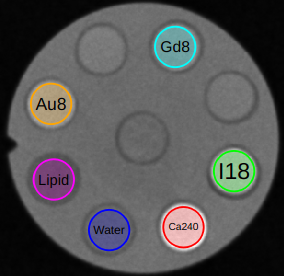}} &
		\subfloat[]{\includegraphics[width=0.48\textwidth]{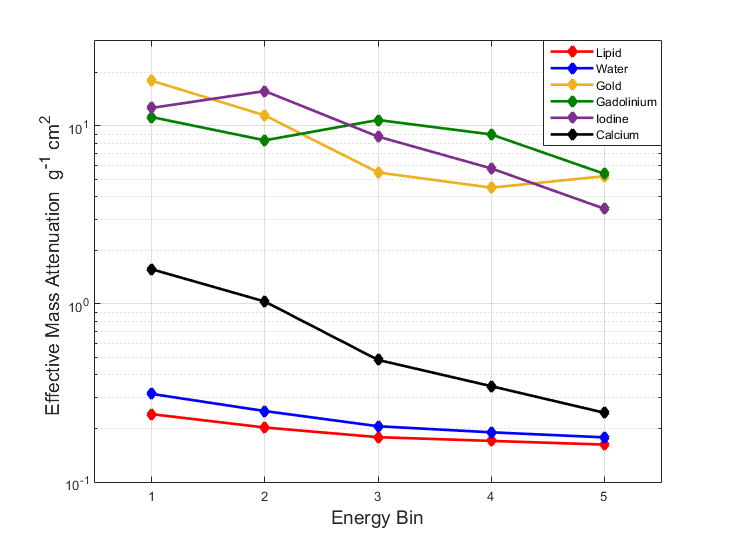}}\\
	\end{tabular}
	\caption{Example of a MARS-MD calibration using reconstructed images of a known phantom; (a) regions covering each material in the reconstructed images are averaged to estimate the effective linear attenuation for each material; (b) from this the effective mass attenuation for each material is estimated and used as the columns in the $M$ matrix to calibrate the MARS-MD algorithm.} 
	\label{fig:Calibration}
\end{figure}

In its implementation within MARS systems, the $M$ matrix in the MARS-MD algorithm is calibrated using scans of phantoms which contain the materials of interest at known concentrations and locations (Fig. \ref{fig:Calibration}). This is a two step process. First the effective linear attenuation for each material (for each concentration and energy bin) is estimated by taking the mean of respective regions in the reconstructed data. These values are then used to calculate the effective mass attenuation, which is used in the $M$ matrix. If more than one concentration is used to calculate the mass attenuation for a given material then it is determined using the non-negative linear least squares estimate of Eqn. \ref{eqn:basis} (where the mass attenuation is taken to be the unknown and concentration taken to be the known variables).

\section{Evaluation}
\subsection{Data Acquisition}
\begin{figure}[h!]
\centering
\includegraphics[width=0.6\textwidth]{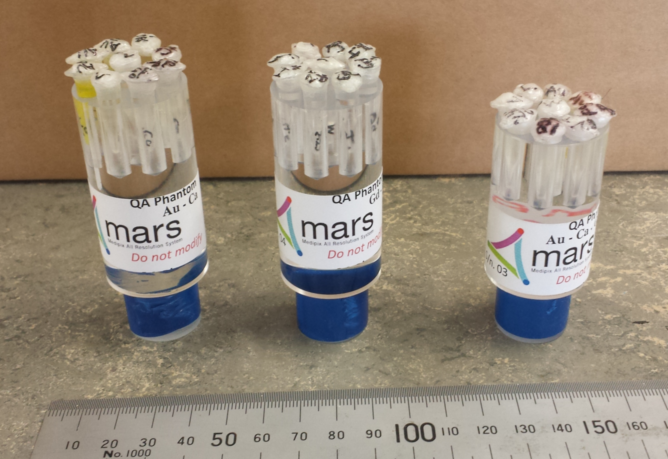}
\caption{CaAu, GdI, and Multi-Contrast QA phantoms (left to right). }
\label{fig:phantoms}
\end{figure}

Scans of the following three phantoms were taken using a version 5 MARS spectral CT scanner. The MARS camera within this scanner contained a single Medipix3RX detector with a $2\,$mm thick CZT sensor layer. The detector was operated in charge summing mode, where the four charge summing counters were set to $15$, $33$, $65$, and $80\,$keV; the three single pixel mode counters were set to $60$, $70$, and $80\,$keV; and the arbitration counter was set just above the noise floor. Only the arbitration and charge summed counters were used for image processing. The x-ray beam was generated by a tungsten anode Source Ray SB-120-350 model x-ray tube operated with a current of $18\,\mu$A and voltage of $120\,$kVp. The x-ray beam was pre-attenuated by $1.8\,$mm aluminium equivalent intrinsic filtration. No extrinsic filtration was used. Scans were taken with a source to object distance of $200\,$mm and a object to detector distance of $50\,$mm. Projection data of each phantom was collected sequentially over five vertical translations of the detector module. 720 open-beam images were collected for each camera position to be used in flat-field normalization. Subtracted energy bins were reconstructed using an in-house iterative reconstruction technique based on ordered subset expectation maximization.         
\newline \newline
The three phantoms that were scanned are known as the CaAu, GdI, and Multi-Contrast QA phantoms (Fig. \ref{fig:phantoms}). Each phantom is made from a $31\,$mm diameter PMMA, with nine $6\,$mm capillaries - eight around periphery and one in the centre. $0.2\,$ml polypropylene Eppendrof tubes containing solutions of materials are inserted into each capillary. Each tube is sealed with paraffin wax. The Multi-Contrast phantom is used for MD calibration and contains vegetable oil; 2x water; $8$ and $2\,$mg(Au)/ml gold chloride; $8$ and $2\,$mg(Gd)/ml diluted MultiHance; $18\,$mg(I)/ml diluted Omnipaque 350; and $240\,$mg(Ca)/ml calcium chloride. The CaAu and GdI phantoms cover the same materials expect at a larger range of concentrations. The CaAu phantom contains water; $8$, $4$, $2$, and $1\,$mg(Au)/ml gold chloride; and $240$, $140$, $70$, and $35\,$mg(Ca)/ml calcium chloride. The GdI phantom contains water; $8$, $4$, $2$, and $1\,$mg(Gd)/ml diluted MultiHance; and $18$, $9$, $4.5$, and $2.25\,$mg(I)/ml diluted Omnipaque 350.  
%
\subsection{Multi-Contrast Phantom}
\begin{figure}[h!]
\centering
\includegraphics[width=0.7\textwidth]{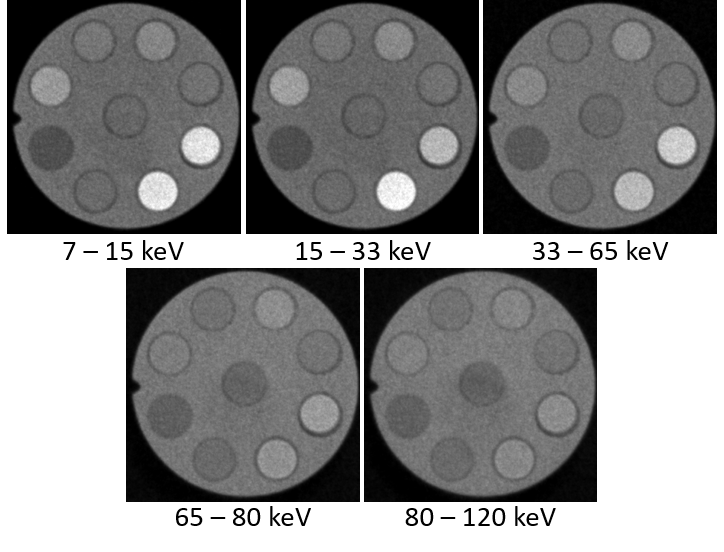}
\caption{Reconstruction of Multi-Contrast phantom with subtracted energy bins displayed with a window of [-700,1000] Hounsfield units. From the indent on the right hand side moving clockwise the capillaries contain $8\,$mg(Au)/ml, $2\,$mg(Au)/ml, $8\,$mg(Gd)/ml, $2\,$mg(Gd)/ml, $18\,$mg(I)/ml, $240\,$mg(Ca)/ml, water, and lipid. The central capillary contains water.}
\label{fig:recon}
\end{figure}

\begin{figure}[t!]
\centering
\includegraphics[width=\textwidth]{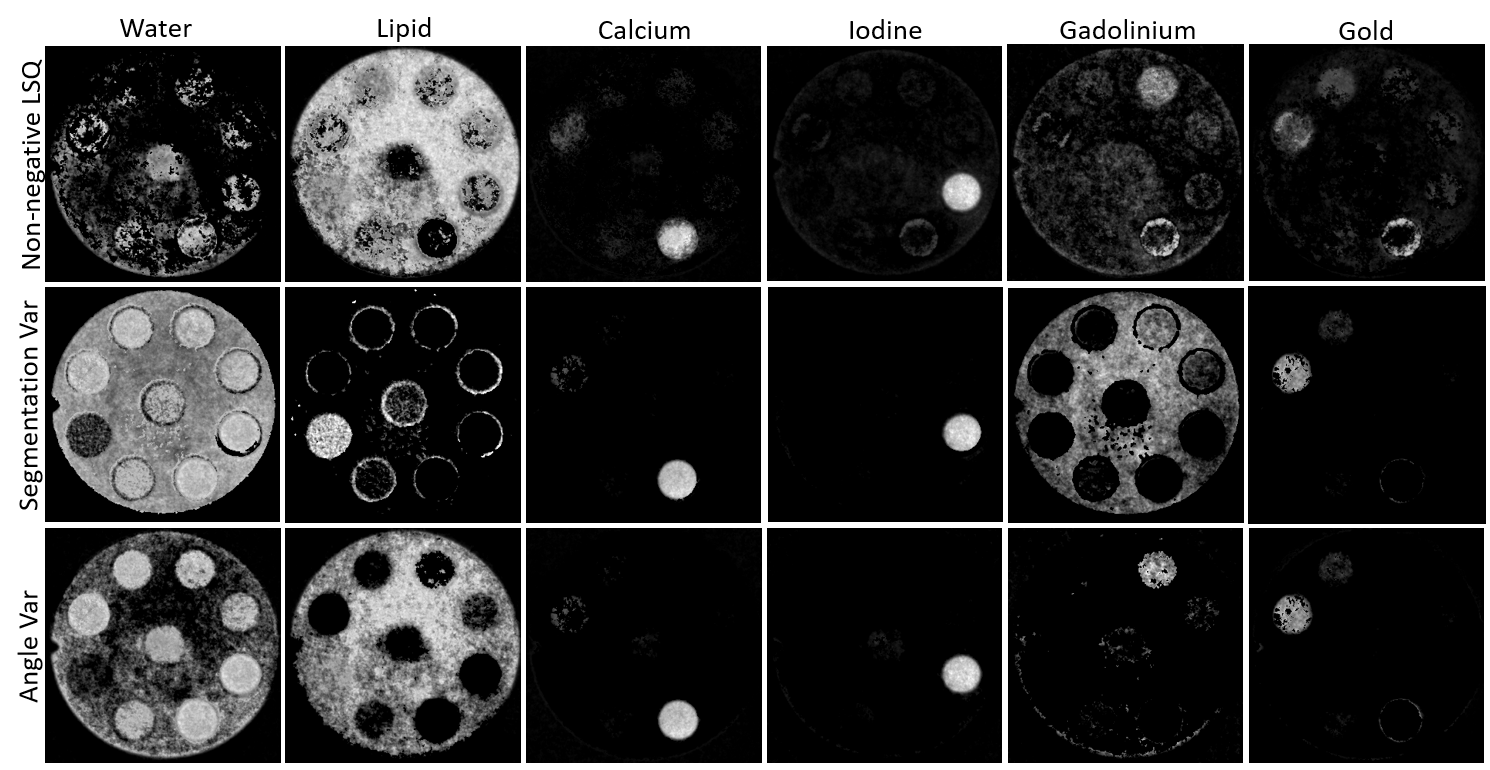}
\caption{Material decomposition of Multi-Contrast Phantom using three different methods - the Angular Rejection and Segmentation variants of MARS-MD, along with an unmodified non-negative linear least squares decomposition. The MARS-MD variants show reasonable separation of the materials contained within each capillary.}
\label{fig:QAMD}
\end{figure}

The reconstruction of a single slice in the middle of the Multi-Contrast phantom is shown in Fig. \ref{fig:recon}. This was processed using three different material decomposition algorithms (Fig. \ref{fig:QAMD}), including the two MARS-MD variants and an unmodified non-negative linear least squares (LSQ). Each MD decomposed five energy bins into six material channels. The MD calibration was calculated from an average 11 slices (including the slice used for MD analysis) with material regions selected to cover the majority of each capillary in a similar way as shown in Fig. \ref{fig:Calibration} (a). 
\newline \newline  
The results from both MARS-MD variants demonstrate reasonable ability to separate the contents of each of the nine capillaries into their appropriate material channels. The exception to this being in the 2mg(Au)/ml and 2mg(Gd)/ml capillaries, where there is difficulty separating low concentration signal from the water background. Due to the enforced sparsity of at most two materials per voxel, there is less crossover of signal between material channels as compared to the non-negative LSQ results. A small amount of degradation of the decompositions are observed in the bottom left quadrant of the phantom, which result from minor beam hardening artefacts in the lower energy bins.     
\newline \newline
The most striking difference between the two MARS-MD variants is how the PMMA signal is split over different material channels. The PMMA has been mostly identified as a combination of gadolinium and water in the Segmentation variant. In this case, the higher signal density of the PMMA has enabled it to be grouped under the dense material category of the segmentation. This makes PMMA appear in the water channel, unlike the other two algorithms where it appears in the lipid channel. The lipid-water volume constraint used in the Segmentation variant also has a tendency of forcing PMMA signal into the water channel under situations where it processed in the soft tissue category. Despite the Angular Rejection variant not describing PMMA with a gadolinium component in this example, we have observed that it can happen when using different data acquisition parameters.

\subsection{Concentration Phantoms}
\begin{figure}[h!]
	\centering
    \begin{tabular}{cc}
		\subfloat[]{\includegraphics[width=0.45\textwidth]{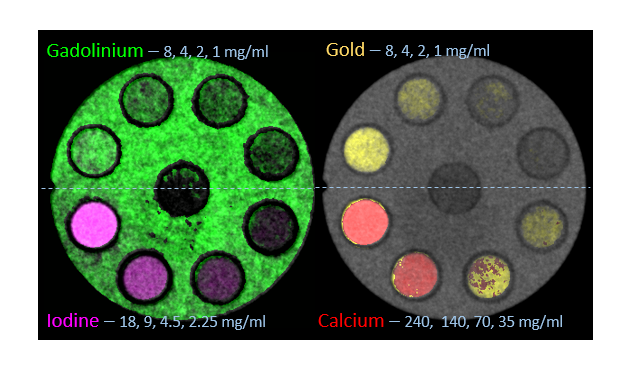}}&
		\subfloat[]{\includegraphics[width=0.45\textwidth]{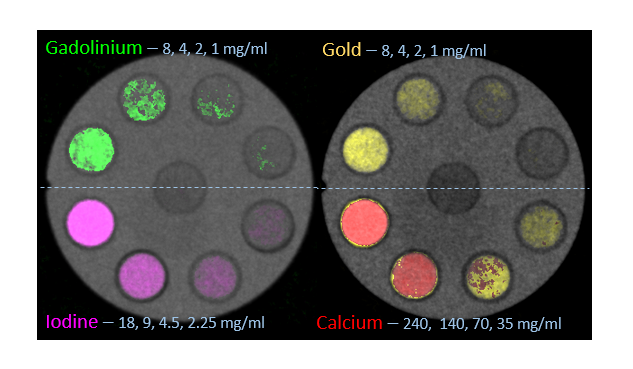}}\\
	\end{tabular}
\caption{Material decomposition of concentration phantoms using (a) Segmentation variant and (b) Angular Rejection variant of MARS-MD showing gold, gadolinium, calcium, and iodine channels fused with the highest energy bin.} 
\label{fig:ConcFusion}
\end{figure}

\begin{figure}[h!]
	\centering
    \includegraphics[width=0.5\textwidth]{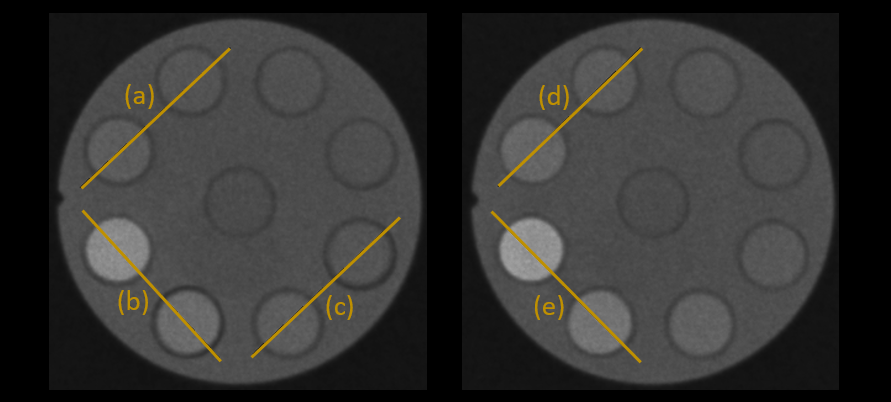}
    \caption{Positions of line profiles displayed in Fig. \ref{fig:lineProfiles} for GdI (left) and CaAu (right) phantoms. Profiles of low concentration capillaries with high levels of misidentification have been excluded. }
    \label{fig:lineProfileScheme}
\end{figure}

\begin{figure}[t!]
	\centering
	\begin{tabular}{ccc}
        \subfloat[]{\includegraphics[width=0.32\textwidth]{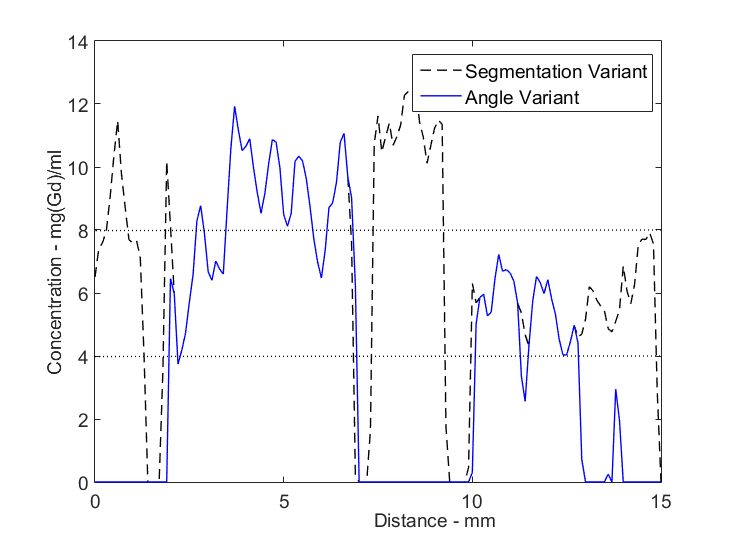}} &
        \subfloat[]{\includegraphics[width=0.32\textwidth]{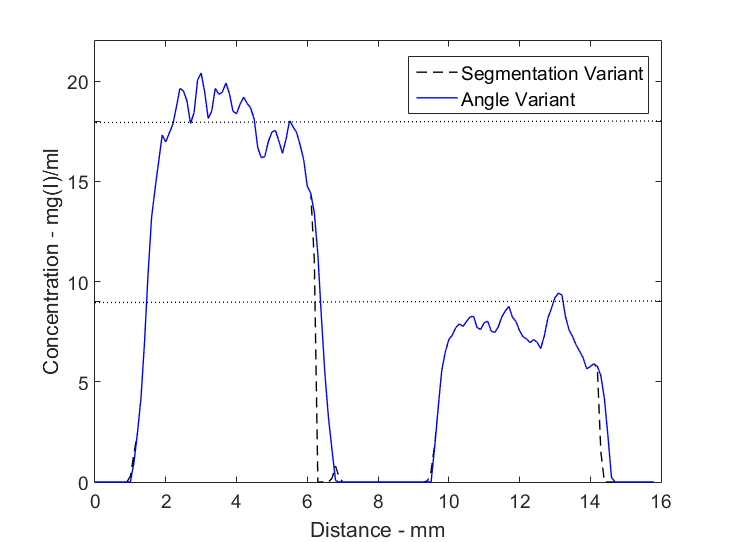}} &
		\subfloat[]{\includegraphics[width=0.32\textwidth]{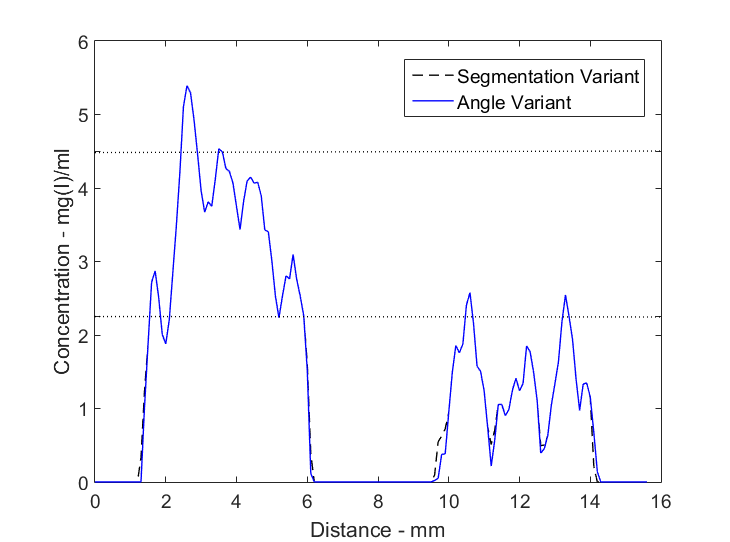}} \\
     \end{tabular}
     \begin{tabular}{cc}
        \subfloat[]{\includegraphics[width=0.32\textwidth]{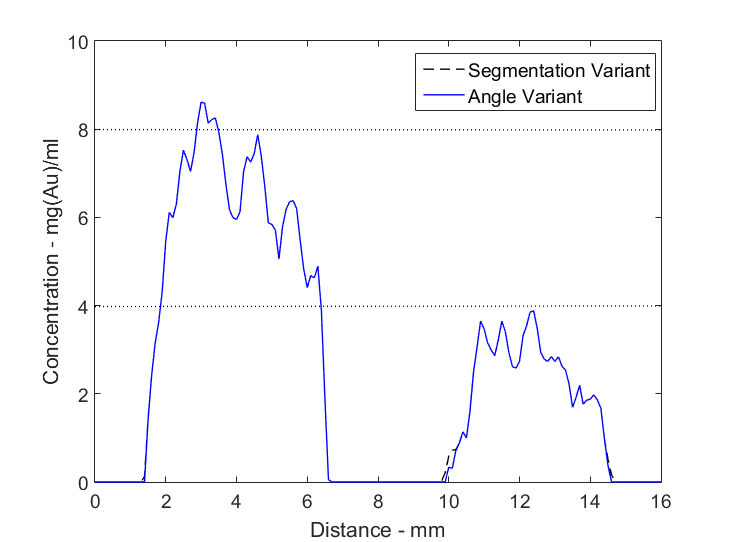}} &
		\subfloat[]{\includegraphics[width=0.32\textwidth]{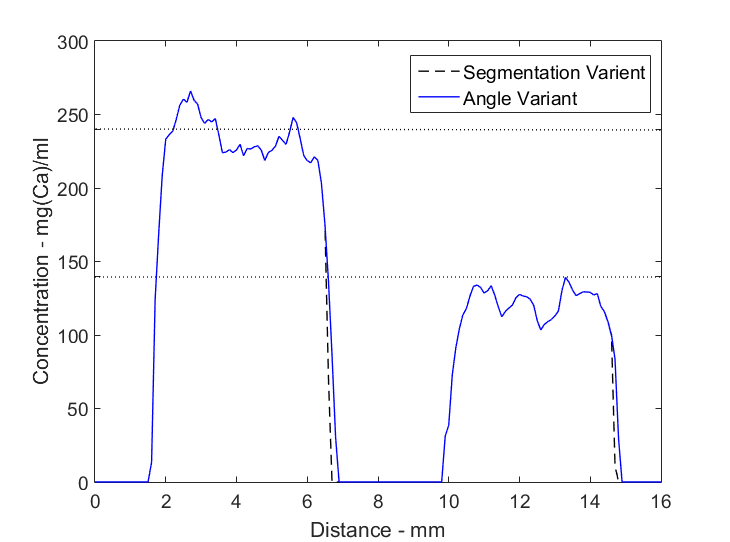}} \\
	\end{tabular}
    \caption{Select line profiles of material decomposed concentration phantoms. Horizontal dotted lines indicate real concentrations of given materials. The profiles cover the following capillaries: (a) $8\,$mg(Gd)/ml and $4\,$mg(Gd)/ml; (b) $18\,$mg(I))/ml and $9\,$mg(I)/ml; (c) $4.5\,$mg(I)/ml and $2.25\,$mg(I)/ml; (d) $8\,$mg(Au)/ml and $4\,$mg(Au)/ml; and (e) $240\,$mg(Ca)/ml and $140\,$mg(Ca)/ml. In this example, most MD results are slightly less than expected yet still close to their desired concentrations.}
    \label{fig:lineProfiles}
\end{figure}

Two phantoms containing four concentrations each of gold, gadolinium, iodine, and calcium solutions were used to assess the quantitative results of the two MARS-MD variants. The concentrations in these phantoms were chosen to cover ranges of both good and poor detectability with the current MARS system. The decomposition of a single slice of these phantoms (excluding the lipid and water channels) is shown in Fig. \ref{fig:ConcFusion}. Line profiles presenting the MD estimated concentrations in selected capillaries is shown in Figs. \ref{fig:lineProfileScheme} and \ref{fig:lineProfiles}. The two lowest concentrations of gadolinium, gold, and calcium have been excluded from the line profiles due to poor identification across the respective capillaries. Due to the way each algorithm has been constructed, the quantitative results for each should be almost equivalent. There are two main exceptions to this. Firstly, the gadolinium distribution is significantly different for reasons stated above. Secondly, the Segmentation variant has a tendency to identify slightly different material boundaries due to the restrictions on voxel materials from its initial segmentation procedure. This is most clearly demonstrated by the $18\,$mg(I)/ml capillary in Fig. \ref{fig:lineProfiles} (b).
\newline \newline
The lowest two concentrations of calcium (Fig \ref{fig:ConcFusion}) are mostly identified as gold. Misidentification such as this is not uncommon at low concentrations. It is also common observe low concentrations of gold misidentified as calcium when using different imaging parameters \cite{Moghiseh2016}. 
\newline \newline
Most of the predicted concentrations shown in the line profile diagrams (Fig. \ref{fig:lineProfiles}) is a small underestimate when compared to the expected concentrations (horizontal dotted lines). Small deviations are expected in image space decomposition for a variety of reasons. For example - despite using narrow energy bins, beam hardening artefacts will still be present. In addition, the extent of the beam hardening will be different in each energy bin. It therefore is not surprising that we have observed that the size of this underestimation changes with modifications to energy bins and x-ray spectrum.         
%
%
%
%
%
%
%

\subsection{Pre-clinical Examples}
Examples of previous publications where the Segmentation variant of the MARS-MD algorithm has been used include: soft tissue imaging using lamb meat \cite{Aamir2014}; arthritic cartilage imaging \cite{Rajendran2017}; intrinsic biomarker and contrast identification in atherosclerosis \cite{Panta2014}; and simultaneous imaging of multiple contrast agents \cite{Moghiseh2016}. Each of these studies use different imaging parameters and provide independent assessments of MD results that are relevant to the given studies. The Angular Rejection variant is more recent and its applications are currently being investigated.


\section{Discussion}
Whereas development of MD techniques in the wider community has largely moved onto iterative joint-reconstruction methods, there is still benefit in further development of image based techniques. Iterative joint-reconstruction techniques are typically very slow when compared to taking the image domain MD route. A more practical approach to the MD problem in spectral CT would be to construct an initial approximation of the decomposed volume using a fast filtered back projection (or similarly fast) reconstruction paired with an image space material decomposition, the results of which are then used for the initial guess in a joint-reconstruction decomposition.
\newline \newline 
The MARS-MD heuristic has a combinatorial approach to solving the MD problem. The typical issues associated with scalability have been minimized by only considering a small subset of material combinations. For the subsets used in the variants above the number of combinations per voxel scales linearly with the number of materials used in the decomposition ($2N-3$ combinations for $N\geq 2$ materials). This approach also provides an easy framework to reject solutions in lieu of others for unique reasons, such as unrealistic material concentrations. 
\newline \newline 
One of the key features of the above MARS-MD algorithms is that they target solutions that are sparse in the material domain. Although this type of constraint has been used in other techniques \cite{Le2011, JWang2011}, it is more common to find techniques which either use conservation constraints or promote spatial uniformity. Strict material sparsity as has been used here has the advantage that it minimizes the risk of fitting materials with K-edges in the imaging range to the noise. It can also turn an under-determined problem (more materials than energy bins) into an over-determined problem. There are drawbacks however. If a region contains more materials than the number allowed by the sparsity condition then it is not possible to correctly identify the contents. The MARS-MD algorithms described above, in this sense, do not facilitate either the identification of multiple high contrast materials within a single voxel nor the combination of high contrast materials and lipid.
\newline \newline
One of the challenges of post-reconstruction MD is the question about how to calibrate them for a wide variety of applications and/or scan samples. Here we have calibrated directly from experimental phantom data. The validity of calibrations for each scan however are dependent on the degree of beam hardening within the sample. This in-turn depends on the size and composition of each scan sample - which will be different. Numerous practices have been followed in within the MARS research team's pre-clinical work to minimize this bias. The phantoms used for calibration are of a similar size and effective density to the samples generally scanned. Sometimes additional calibration vials have been included alongside initial samples to test the validity of the MD calibration and to correct it if necessary.        
\newline \newline
Lastly, the quality of results produced by any MD technique is highly dependent on the information measured from the spectral domain. What counts as good or sufficient spectral information can depend on things such as the decomposition bases that are used; what and how many energy bins are used; and even the degree of beam hardening caused by the sample. In other words, a set of optimal acquisition parameter for one application may not be optimal for other applications. We have given reference to a diverse range of applications using our methodology for this reason. This problem of spectral CT scan protocol optimality is currently an area of active research.

\section{Conclusion}
This paper has described two constrained image space material decomposition techniques that have been used in MARS spectral CT systems over the past several years. This description includes an outline of the heuristic that the MARS-MD algorithms have been constructed from; the commonly used Segmentation variant algorithm; and the more recent Angular Rejection variant algorithm. A brief example of their application to spectral CT scans of generic multiple-contrast phantoms demonstrates reasonable material identification performance. 




\acknowledgments





\bibliographystyle{ieeetr}
\bibliography{references}







\end{document}